\documentclass[
   preprint,
superscriptaddress,
showpacs,
 amsmath,amssymb,
 aps,
 pre,
 nobibnotes,
]{revtex4}
\usepackage{color,soul}

\usepackage{graphicx}
\usepackage{subfig}
\usepackage{booktabs}
\usepackage{dcolumn}
\usepackage{bm}
\usepackage[mathlines]{lineno}
\usepackage{xspace}
\usepackage{url}

\begin{document}
\graphicspath{./images}

\title{Vascular networks due to dynamically arrested crystalline ordering of elongated cells}
\author{Margriet M. Palm}
\email{M.M.Palm@cwi.nl}
\affiliation{Centrum Wiskunde \& Informatica, Amsterdam, The Netherlands}
\affiliation{Netherlands Consortium for Systems Biology - Netherlands Institute for Systems Biology, Amsterdam, The Netherlands}
\author{Roeland M. H. Merks}
\email{merks@cwi.nl}
\affiliation{Centrum Wiskunde \& Informatica, Amsterdam, The Netherlands}
\affiliation{Netherlands Consortium for Systems Biology - Netherlands Institute for Systems Biology, Amsterdam, The Netherlands}
\affiliation{Mathematical Institute, Leiden University, Leiden, The Netherlands}
\date{\today}

\begin{abstract}
Recent experimental and theoretical studies suggest that crystallization and glass-like solidification are useful analogies for understanding cell ordering in confluent biological tissues. It remains unexplored how cellular ordering contributes to pattern formation during morphogenesis. With a computational model we show that a system of elongated, cohering biological cells can get dynamically arrested in a network pattern. Our model provides a new explanation for the formation of cellular networks in culture systems that exclude intercellular interaction via chemotaxis or mechanical traction.
\end{abstract}

\maketitle
\section{Introduction}
By aligning locally with one another, cells of elongated shape form ordered, crystalline configurations in cell cultures of, e.g. fibroblasts \cite{Elsdale1968,pietak2008}, mesenchymal stem cells \cite{pietak2008},
and endothelial cells \cite{Szabo:2010p7067}. Initially the cells form
small clusters of aligned cells; the clusters then grow and the range
over which cells align increases with time
\cite{Elsdale1976,pietak2008}. To study the emergence of such
crystalline cellular ordering, it is useful to make an analogy with
liquid crystals \cite{pietak2008}. For example, a ``cellular
temperature'' can be defined to describe the cell-type specific
persistence (low cellular temperature) or randomness (high cellular
temperature) of cell motility, where cells of high cellular
temperature ({\em e.g.,} fibroblasts) are less likely to form
crystalline configurations than cells of low temperature ({\em e.g.},
mesenchymal stem cells) \cite{pietak2008}. It was similarly proposed
that collective cell motion in crowded cell sheets can be
understood as system approaching a glass transition
\cite{Angelini2011,Garrahan2011}.
Although these studies provide useful insights into the ordering of
cells in confluent cell layers, it remains unexplored how
crystallization and glass-like dynamics contribute to the formation of
more complex shapes and patterns during biological morphogenesis.

Cells' organizing into network-like structures, as it occurs for
example during blood vessel development, is a suitable system to study
how cellular ordering participates in pattern formation.
In cell cultures after stimulation by growth factors (VEGFs, FGFs),
endothelial cells elongate and form vascular-like network
structures~\cite{Cao1998,Drake2000,Parsa22032011}. The mechanisms that
drive the aggregation of endothelial cells and their subsequent
organization into network is a subject of debate. Most models assume
an attractive force between cells, either due to chemotaxis \cite{Gamba2003,Serini2003,Ambrosi2004,Merks2004,Merks2006b,Merks2006a,Merks2008,Kohn-Luque2011,Scianna2011}
or due to mechanical traction via the
extracellular matrix \cite{Manoussaki1996,Manoussaki2002,Murray2003,Tranqui2000,Namy2004a,Tracqui2002}.
\emph{In vitro} experiments show that astroglia-related rat C6 cells and
muscle-related C212 cells can form network-like structures on a rigid
culture substrate \cite{Szabo2007}, which excludes formation of
mechanical or chemical attraction between cells.  Therefore a second
class of explanations proposed that cells form networks by adhering
better to locally elongated configurations of cells \cite{Szabo2007}
or elongated cells \cite{Szabo2008}. Here we show that, in absence of
mechanical or chemical fields such mechanisms are unnecessary:
elongated cells organize into network structures if they move and
rotate randomly, and adhere to adjacent cells. As the cells align
locally with one another, a network pattern appears.
Additional, long-range cell-cell attraction mechanisms, e.g., chemotaxis or mechanotaxis, act to
stabilize the pattern and fix its wave length.

\begin{figure}[hbtp]
 \centering
 \includegraphics[width=\columnwidth]{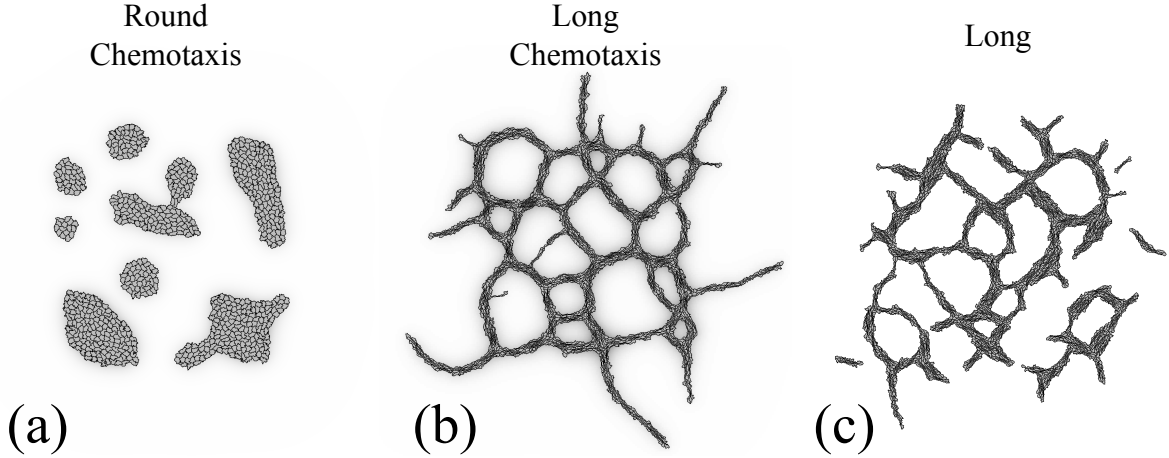}
  \caption{Effect of chemotaxis and cell shape on pattern formation. \textbf{A} round, chemotacting, and adhesive cells (10,000 MCS), \textbf{B} elongated, chemotacting and adhesive cells (10,000 MCS), and \textbf{C} elongated, non-chemotacting and adhesive cells (250,000 MCS). In all panels 700 cells are seeded on the center 500x500 pixels of an 800x800 lattice.}
\label{fig:models}
\end{figure}

\section{Model description}
To model the collective movement of elongated cells, we use the cellular Potts method (CPM), {\em aka} the Glazier-Graner-Hogeweg model~\cite{Graner1992,Glazier1993a}, a lattice-based, Monte-Carlo model that has been used to model developmental mechanisms including somitogenesis \cite{Glazier2008,Hester2011}, convergent extension \cite{Zajac2000a} and fruit fly retinal patterning \cite{Kafer2007}. The CPM represents cells as connected patches of lattice sites with identical spin $\sigma\in\mathbb{N}$; lattice sites with spin $\sigma=0$ represent the extracellular matrix (ECM). To simulate stochastic cell motility, the CPM iteratively displaces cell-cell and cell-ECM boundaries by attempting to copy the spin of a randomly selected site into a randomly selected adjacent lattice site $\vec{x}$, monitoring the resulting change $\Delta H$ of a Hamiltonian,

\begin{align}
    H =&  \sum_{\left(\vec x,\vec{x}^\prime\right)}J(\sigma(\vec x),\sigma(\vec{x}^\prime))\left(1-\delta(\sigma(\vec x),\sigma(\vec{x}^\prime))\right)\nonumber \\
    & + \sum_{\sigma}  \lambda_\text{A} \left ( a(\sigma) - A \right )^2 +  \sum_{\sigma}  \lambda_\text{L} \left (l(\sigma) - L \right )^2 . \label{eq:ham}
\end{align}

\noindent A copy attempt will always be accepted if $\Delta H\leq0$, if $\Delta H>0$ a copy attempt is accepted with the Boltzmann probability $P(\Delta H) = \exp(-{\Delta H}/{\mu(\sigma)})$, with $\mu(\sigma)$ a ``cellular temperature'' to simulate cell-autonomous random motility. For simplicity, we here assume that all cells have identical temperature. The time unit is a Monte Carlo step (MCS), which corresponds with as many copy attempts as there are lattice sites.

The first term of Eq. \ref{eq:ham} defines an adhesion energy, with the Kronecker delta returning a value of 1 for site pairs at cell-cell and cell-ECM interfaces, or zero otherwise. In the model two contact energies are defined: $J_\text{cell,cell}$ for $\sigma>0$ at both lattice sites, and $J_\text{cell,ECM}$ for $\sigma=0$ at one lattice site. The second and third term are shape constraints that penalize deviations from a target shape, with $A$ and $L$ a target area and length, and $a(\sigma)$ and $l(\sigma)$ the current area and length of the cell; $\lambda_\text{A}$ and $\lambda_\text{L}$ are shape parameters. We efficiently estimate $l(\sigma)$ by keeping track of a cellular inertia tensor as previously described \cite{Merks2006b}.

In a subset of simulations, we further assume that cells secrete a diffusing  chemoattractant $c$, which we describe with a partial differential equation:
\begin{equation}
{\frac{\partial c(\vec x,t)}{\partial t}} = D\nabla^2c(\vec x,t) + s(1-\delta(\sigma(\vec{x}),0)) - \epsilon\;\delta(\sigma(\vec{x}),0))  ,\label{eq:field}
\end{equation}

\noindent with diffusion constant $D$, secretion rate $s$ and decay
rate $\epsilon$.  After each MCS, a forward Euler method solves
Eq.~\ref{eq:field} for 15 steps with $\Delta t=2\;\text{s}$ with zero
boundary conditions. To model the cells' chemotaxis up concentration
gradients of the chemoattractant, during each copy attempt from $\vec{x}$ to $\vec{x^\prime}$ we increase
$\Delta H$ with a $\Delta H_\text{chemotaxis} = \lambda_c
\left(c(\vec{x})-c(\vec{x}^\prime)\right)$, with $\lambda_c$ a
chemotactic strength \cite{Savill1996}. One lattice unit (l.u.)
corresponds with $2\;\mathrm{\mu m}$. We use the following parameter settings,
unless specified otherwise: $\mu = 1$; $J_\text{cell,cell}$ = .5;
$J_\text{cell,ECM}$ = .35; $\lambda_\text{A} = 1$; $\lambda_\text{L} =
.1$; $\lambda_c = 10$; $A = 100$ l.u.$^2$; $L = 60$ l.u.; $D =
10^{-13}\;\text{m}^2 \text{s}^{-1}$; $\epsilon = 1.8\cdot10^{-4}
\;\text{s}^{-1}$; $s = 1.8\cdot10^{-4}\;\text{s}^{-1}$. Unless stated
otherwise, a simulation is initialized with 175 cells randomly
distributed on a 220x220 area at the center of a 400x400 lattice.

\section{Results}
\begin{figure}[t]
 \centering
  \includegraphics{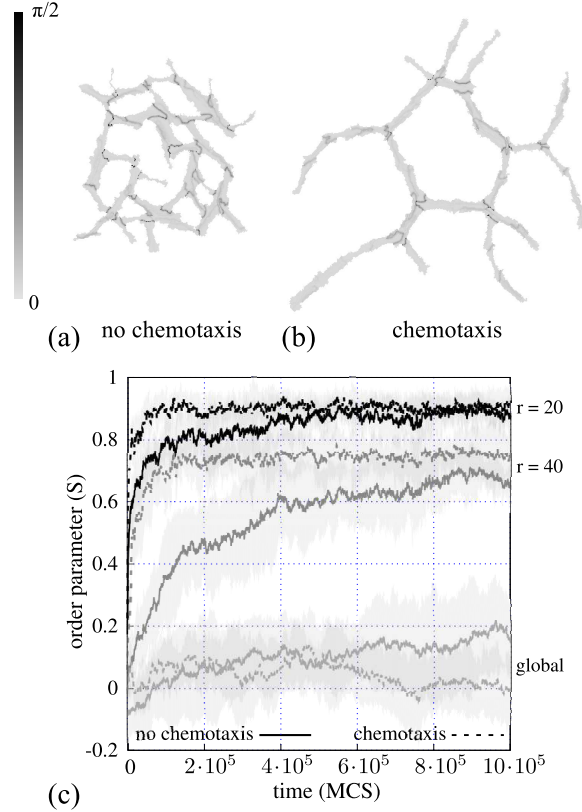}
  \caption{Crystalline cell ordering during network formation.
    \textbf{A}-\textbf{B} $\theta(\vec{x},r)$ with $r=3$ for a
    simulation with chemotaxis (\textbf{A}) and without chemotaxis (\textbf{B})
    after 25,000 MCS. \textbf{C} Temporal evolution of orientational
    order parameter $S(r)$ for $r=20$ (black curves), $r=40$ (gray
    curves) and $r\rightarrow\infty$ (light gray) without chemotaxis
    (solid) and with chemotaxis (dashed). Order parameter is averaged
    over 10 simulation repeats (gray shadows represent standard
    deviation).}
 \label{fig:op}
\end{figure}

As Fig.~\ref{fig:models} shows, and in agreement with previous reports
\cite{Merks2006b}, if we allow for chemotaxis, rounded cells
accumulate into rounded clusters (Fig.~\ref{fig:models}\textbf{A}) and
elongated cells aggregate into networks (Fig.~\ref{fig:models}\textbf{B}).
Interestingly, however, chemotaxis is not required for network
formation: cell-cell adhesion between elongated cells suffices for
forming networks (Fig.~\ref{fig:models}\textbf{B}). Movies corresponding with
Fig.~\ref{fig:models}\textbf{B} and \textbf{C} \footnote{See Supplemental Material at [URL will
  be inserted by publisher] for Movie S1, model without chemotaxis,
  and Movie S2, model with chemotaxis} suggest that the gradual
alignment of cells with their neighbors is key to network formation
and network evolution. To characterize this cell alignment, we define
$\theta(\vec{x},r)$ as the angle between the direction of the long
axis $\vec{v}(\sigma(\vec{x}))$ of the cell at $\vec{x}$, and a local
director $\vec{n}(\vec{x},r)$, a weighted local average of cell
orientations defined at radius $r$ around $\vec{x}$:
$\vec{n}(\vec{x},r)=\left\langle\vec{v}(\sigma(\vec{y}))\right\rangle_{\left\{\vec{y}\in\mathbb{Z}^2:\left|\vec{x}-\vec{y}\right|<r\right\}}$. Figure
~\ref{fig:op}\textbf{A} and \textbf{B} depict the value of $\theta(\vec{x},3)$ for simulations
without chemotaxis (Fig. ~\ref{fig:op}\textbf{A}) and with chemotaxis (Fig. ~\ref{fig:op}\textbf{B}), with dark gray
values indicating values of $\theta(\vec{x},3) \to \pi/2$. Network
branches are separated by large values of $\theta(\vec{x},3)$,
indicating that within branches cells are aligned, whereas branch
points are ``lattice defects'' in which cells with different
orientations meet.

Supplemental Movies S3 and S4 \footnote{See Supplemental Material at [URL will be inserted by publisher] for Movie S3,  $\theta(\vec{x},3)$ without chemotaxis, and Movie S4, $\theta(\vec{x},3)$ with chemotaxis} show how the cells align gradually over time in the absence and presence of chemotaxis. To characterize the temporal development of cell alignment in more detail, we use an orientational order parameter
$S(r)=\left \langle \cos(2 \theta(\vec{X}(\sigma),r)) \right\rangle_\sigma$ \cite{DeGennes1993}
with $\vec{X}(\sigma)$ the center of mass of cell $\sigma$. $S$ ranges from 0 for randomly oriented cells to 1 for cells oriented in parallel.

Figure \ref{fig:op}\textbf{C} shows the evolution of the global
orientational order parameter $\lim_{r\to\infty}S(r)$ and of the local
orientational order parameters $S(20)$ and $S(40)$. Both with
chemotaxis (dashed lines) and without (solid lines), $S(20)$ grows
more quickly and reaches higher ordering than $S(40)$. The reason for this is that in cells of length
$50-60$ l.u., $S(20)$ (covering cells up to a radius $r=20$ from the
cell's center of mass) only detects lateral alignment of cells,
whereas a radius $S(40)$ also detects linear line-up of cells. Thus
cell-cell adhesion of long cells quickly aligns cells with the left
and right neighbors, while it aligns them more slowly with those in
front and behind. This results in networks with short branches of aligned
cells. Interestingly, chemotaxis aligns cells more rapidly, both along the short and long sides of cells, resulting in networks with much longer branches than with adhesion alone.

\begin{figure}[hbt]
  \centering
  \includegraphics[width=\linewidth]{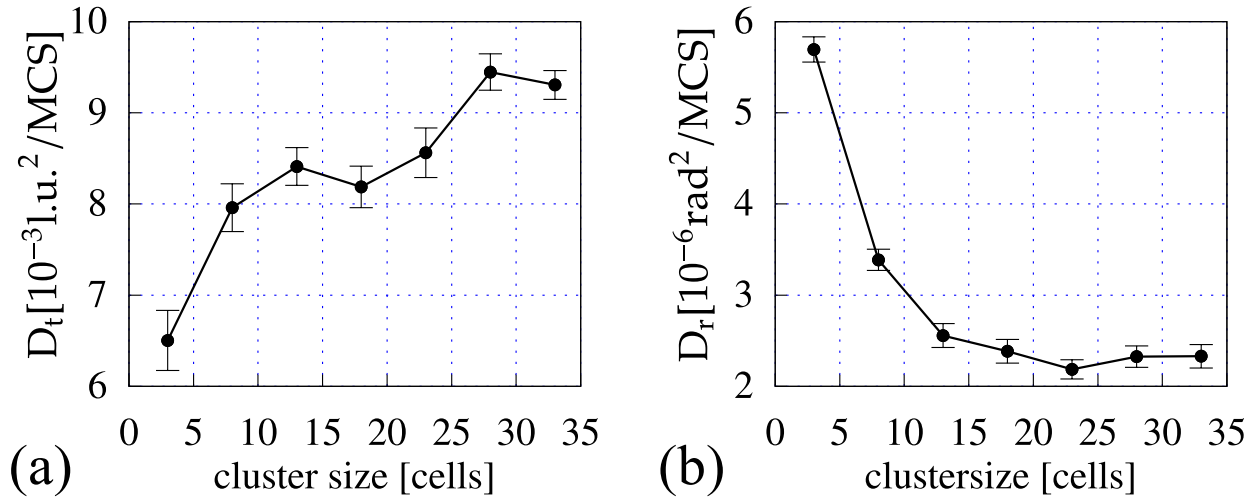}
  \caption{Relation between cluster size and cell displacement.
Clusters
    are calculated for each morphology between 500 and 25,0000 MCS (100
    simulation repeats), with an interval of 500 MCS; see text for details.
The error bars represent the standard error of the linear fits used to estimate diffusion coefficients.}
  \label{fig:cluster}
\end{figure}

Next we analyze the mechanisms that drive the orientational ordering
in the cell networks. Visual inspection of the simulation movies
suggests that single cells move and rotate much more rapidly than
locally aligned clusters of cells. A network of locally aligned cells
forms rapidly from initially dispersed cells. Merging of branches
seems to be a much slower process, and potentially prevents a further
evolution to global nematic order. To quantify these observations we
measured the translational and rotational diffusion coefficients of
cells as a function of the size of the network branch to which it belongs.
We loosely define a network branch, or
{\em cluster} of aligned cells as a connected set of at least two cells with relative orientations
$<5^{\circ}$, {\em i.e.}, in Fig.~\ref{fig:op}\textbf{A} and \textbf{B} dark gray values separate
the clusters.  To detect clusters computationally, we first identify the
connected sets for which $\theta(\vec{x},3)\leq 5^{\circ}$, which are
surrounded by lattice sites of $\sigma=0$ or sites with
$\theta(\vec{x},3)>5^{\circ}$. We then eliminate connected sets of fewer
than fifty lattice sites. The CPM cells sharing at least 50\% of their
lattice sites with one of the remaining sets form a cluster.
The translational diffusion coefficient, $D_\text{t}$,
derives from the mean square displacement (MSD) of a set of cells:
$\left \langle| \vec{X}(\sigma,t)-\vec{X}(\sigma,0)|
  ^2\right \rangle_{\sigma} = 4 D_\text{t}t$. Similarly, the
rotational diffusion coefficient, $D_\text{r}$, derives from the mean
square rotation (MSR) of a set of cells: $\langle
\left ( \alpha(\sigma,t)-\alpha(\sigma,0) \right )^2 \rangle_{\sigma}
= 2 D_\text{r}t$, with $\alpha(\sigma,t)-\alpha(\sigma,0)$ the angular
displacement of a cell between time 0 and $t$.
During a simulation, cells may move between clusters, and
clusters can merge. Therefore, to calculate $D_\text{t}$ and $D_\text{r}$ of cells
as a function of cluster size, for 100 simulations of 250,000 MCS we
measured trajectories of each individual cell with one data point per 500 MCS, and kept track of the
size of the cluster it was classified into at each time point. We defined cluster size bins, with the first bin collecting all clusters consisting of two to five cells, and the next bins running from 6 to 10, 11 to 15, {\em etc.}  We split up the trajectories into chunks of 10 consecutive data points, during which the cells stayed within clusters belonging to one bin. To calculate $D_\text{t}$ and $D_\text{r}$ we performed a least square fitting on the binned MSD and MSR values for these trajectory chunks.


The translational diffusion, $D_\text{t}$, increases slightly with
cluster size (Fig. \ref{fig:cluster}\textbf{A}). This may reflect that
the probability of hopping between small clusters will be larger than the
probability of hopping between larger clusters, resulting in an
overrepresentation of slow cells in the small clusters. Interestingly,
the rotational diffusion $D_\text{r}$ drops with the cluster size
(Fig. \ref{fig:cluster}\textbf{B}), indicating that cells in large
clusters rotate more slowly. These results suggest that
the rotation of cells in big clusters is limited, which
reduces the probability that two clusters rotate and merge into a
single larger cluster. Therefore, if the size of clusters increases,
their rotation speeds drop as does the probability of cluster fusion.
Thus, although further alignment of clusters would reduce the pattern energy
$H$ (Eq.~\ref{eq:ham}), the pattern evolution essentially freezes.

\begin{figure}
 \centering
 \includegraphics[width=\linewidth]{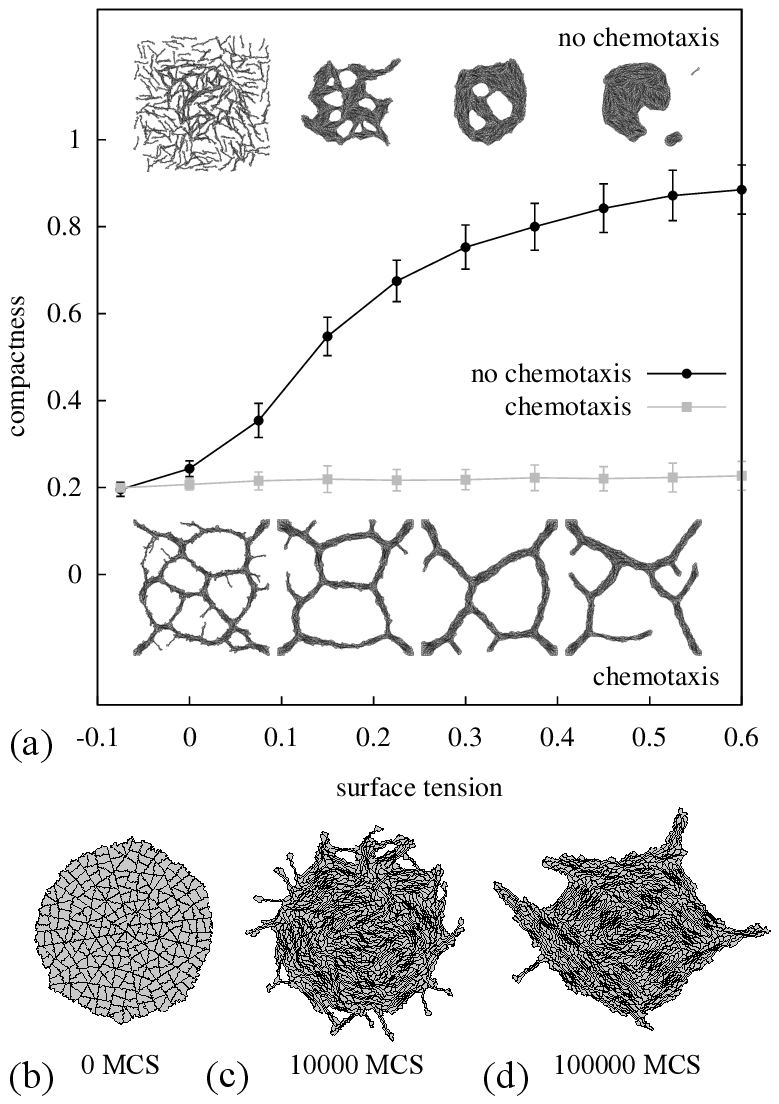}
 \caption{\textbf{A} Relation between compactness and surface tension, with and without chemotaxis. The compactness was calculated at 100,000 MCS and averaged over 100 simulations (error bars represent standard deviation). Simulations were initialized with 350 cells on 260x260 path on the center of a 420x420 lattice. \textbf{B}-\textbf{D} evolution of a simulation initialized with 128 cell blob on the center of a 420x420 grid.}
 \label{fig:parrun}
\end{figure}

To corroborate our hypothesis that network patterns are transient
patterns that increasingly slowly evolve towards nematic order, we
looked for model parameters that could speed up pattern evolution.
Fig.~\ref{fig:parrun}\textbf{A} shows the effect of surface tension ($\gamma_\text{cell,ECM}$) on
the ability of cells to form networks after 100,000 MCS, as expressed
by the configuration's {\em compactness} $C =
\tfrac{A_\text{cells}}{A_\text{hull}}$, where  $A_\text{hull}$ is the area
of the convex hull of the largest connected group of cells,
and $A_\text{cells}$ is the summed area of the cells inside the hull.
A value of $C\rightarrow 1$ indicates a spheroid of cells, where for networks $C$ would tend to zero.
For values of $\gamma_\text{cell,ECM} = J_\text{cell,ECM} -
\frac{J_\text{cell,cell}}{2} > 0$, the equilibrium pattern should
minimize its surface area with the ECM. Indeed at increased surface
tensions the cells settle down in spheroids or networks with only few
meshes, although they initially still form network-like patterns (see Movie
S5 \footnote{See Supplemental Material at [URL Will be inserted by
publisher] for Movie S5, model without chemotaxis and $\gamma_\text{cell,ECM}=0.3$}).
To confirm that also for $\gamma_\text{cell,ECM}=0.1$ ({\em i.e.}, the
values used in Figs.~1-3) spheroids are stable configurations, we
initialized our model with a spheroid (Fig.~\ref{fig:parrun}\textbf{B}). Although initially some cells sprout
(Fig.~\ref{fig:parrun}\textbf{C}) from the spheroid due to their
elongation, they then align gradually and the cell cluster remains spherical. No network formation was detected in
simulations of 100,000 MCS (Fig.  \ref{fig:parrun}\textbf{D}),
suggesting that spheroids represent the global minimum of the
Hamiltonian. Interestingly, in presence of chemotaxis networks form
for a wide range of surface tensions (inset Fig.~\ref{fig:parrun}\textbf{A}
and~\cite{Merks2006b}).

\section{Discussion}
Our analysis suggests that in the cellular Potts model elongated,
adhesive cells can form networks in a parameter regime where a
spheroid pattern is the minimal energy state. The cells initially
align with nearby cells, thus forming the branches of the network. In
order for the pattern to evolve further towards the minimal-energy
spheroid pattern, the locally aligned clusters of cells must join
adjacent branches, for which they must move and rotate. Our analysis
of the rotational and translational diffusion of cells in
Fig.~\ref{fig:cluster} shows that this becomes more difficult for
cells belonging to larger clusters. Thus the networks evolve ever more
slowly to the minimal energy state, and gets dynamically arrested in a
network-like configuration, a phenomenon reminiscent of the glass
transition, as e.g. observed in attractive colloid systems
\cite{Foffi2005}, collective cell migration of biological cells {\em in vitro}
\cite{Angelini2011,Garrahan2011}, and colloid rod suspensions \cite{Solomon2010} in which gels can form from clusters of parallel rods \cite{Bernal:1941to,vanBruggen:2002hf,Wilkins:2009ht}.

Fig.~\ref{fig:parrun}\textbf{A} suggests that the cellular Potts simulations
undergo a glass transition as the surface tension drops: for high
surface tension the system evolves towards equilibrium, for lower
surface tensions the system becomes jammed in a network-like state.
Thus our model provides a new explanation for the formation of
vascular networks in absence of chemical or mechanical, long-range, intercellular
attraction \cite{Szabo2007}. Interestingly, intercellular attraction via chemotaxis stabilizes the formation of networks in our simulations \cite{Merks2006b} and can drive sprouting from spheroids (not shown). This
suggests that networks are an equilibrium pattern of our system in
presence of intercellular attraction.  Nevertheless the present
analysis of arrested dynamics provides new insight into the system
with intercellular attraction: chemotaxis reinforces local ordering
over a distance proportional to the diffusion length of the
chemoattractant producing networks of a scale independent of surface
tension \cite{Merks2006b}.

\begin{acknowledgments}
  The authors thank the IU and the Biocomplexity Institute for
  providing the CC3D modeling environment (\url{www.compucell3d.org})
  \cite{Swat2012} and SARA for providing access to The National
  Compute Cluster LISA (\url{www.sara.nl}).  This work was
  financed by the Netherlands Consortium for Systems Biology
  (NCSB) which is part of the Netherlands Genomics
  Initiative/Netherlands.  The investigations were in part supported
  by the Division for Earth and Life Sciences (ALW) with financial aid
  from the Netherlands Organization for Scientific Research (NWO).

\end{acknowledgments}


\begin{thebibliography}{40}
\expandafter\ifx\csname natexlab\endcsname\relax\def\natexlab#1{#1}\fi
\expandafter\ifx\csname bibnamefont\endcsname\relax
  \def\bibnamefont#1{#1}\fi
\expandafter\ifx\csname bibfnamefont\endcsname\relax
  \def\bibfnamefont#1{#1}\fi
\expandafter\ifx\csname citenamefont\endcsname\relax
  \def\citenamefont#1{#1}\fi
\expandafter\ifx\csname url\endcsname\relax
  \def\url#1{\texttt{#1}}\fi
\expandafter\ifx\csname urlprefix\endcsname\relax\def\urlprefix{URL }\fi
\providecommand{\bibinfo}[2]{#2}
\providecommand{\eprint}[2][]{\url{#2}}

\bibitem[{\citenamefont{Elsdale}(1968)}]{Elsdale1968}
\bibinfo{author}{\bibfnamefont{T.~R.} \bibnamefont{Elsdale}},
  \bibinfo{journal}{Exp. Cell Res.} \textbf{\bibinfo{volume}{51}},
  \bibinfo{pages}{439} (\bibinfo{year}{1968}).

\bibitem[{\citenamefont{Pietak and Waldman}(2008)}]{pietak2008}
\bibinfo{author}{\bibfnamefont{A.}~\bibnamefont{Pietak}} \bibnamefont{and}
  \bibinfo{author}{\bibfnamefont{S.~D.} \bibnamefont{Waldman}},
  \bibinfo{journal}{Phys. Biol.} \textbf{\bibinfo{volume}{5}},
  \bibinfo{pages}{016007} (\bibinfo{year}{2008}).

\bibitem[{\citenamefont{Szab{\'o} et~al.}(2010)\citenamefont{Szab{\'o}, Unnep,
  M{\'e}hes, Twal, Argraves, Cao, and Czir{\'o}k}}]{Szabo:2010p7067}
\bibinfo{author}{\bibfnamefont{A.}~\bibnamefont{Szab{\'o}}},
  \bibinfo{author}{\bibfnamefont{R.}~\bibnamefont{Unnep}},
  \bibinfo{author}{\bibfnamefont{E.}~\bibnamefont{M{\'e}hes}},
  \bibinfo{author}{\bibfnamefont{W.~O.} \bibnamefont{Twal}},
  \bibinfo{author}{\bibfnamefont{W.~S.} \bibnamefont{Argraves}},
  \bibinfo{author}{\bibfnamefont{Y.}~\bibnamefont{Cao}}, \bibnamefont{and}
  \bibinfo{author}{\bibfnamefont{A.}~\bibnamefont{Czir{\'o}k}},
  \bibinfo{journal}{Phys. Biol.} \textbf{\bibinfo{volume}{7}},
  \bibinfo{pages}{046007} (\bibinfo{year}{2010}).

\bibitem[{\citenamefont{Elsdale and Wasoff}(1976)}]{Elsdale1976}
\bibinfo{author}{\bibfnamefont{T.}~\bibnamefont{Elsdale}} \bibnamefont{and}
  \bibinfo{author}{\bibfnamefont{F.}~\bibnamefont{Wasoff}},
  \bibinfo{journal}{Dev. Genes Evol.} \textbf{\bibinfo{volume}{147}},
  \bibinfo{pages}{121} (\bibinfo{year}{1976}).

\bibitem[{\citenamefont{Angelini et~al.}(2011)\citenamefont{Angelini, Hannezo,
  Trepat, Marquez, Fredberg, and Weitz}}]{Angelini2011}
\bibinfo{author}{\bibfnamefont{T.~E.} \bibnamefont{Angelini}},
  \bibinfo{author}{\bibfnamefont{E.}~\bibnamefont{Hannezo}},
  \bibinfo{author}{\bibfnamefont{X.}~\bibnamefont{Trepat}},
  \bibinfo{author}{\bibfnamefont{M.}~\bibnamefont{Marquez}},
  \bibinfo{author}{\bibfnamefont{J.~J.} \bibnamefont{Fredberg}},
  \bibnamefont{and} \bibinfo{author}{\bibfnamefont{D.~A.} \bibnamefont{Weitz}},
  \bibinfo{journal}{P. Natl. Acad. Sci. U.S.A.} pp. \bibinfo{pages}{1--6}
  (\bibinfo{year}{2011}).

\bibitem[{\citenamefont{Garrahan}(2011)}]{Garrahan2011}
\bibinfo{author}{\bibfnamefont{J.~P.} \bibnamefont{Garrahan}},
  \bibinfo{journal}{P. Natl. Acad. Sci. U.S.A.} \textbf{\bibinfo{volume}{108}},
  \bibinfo{pages}{4701} (\bibinfo{year}{2011}).

\bibitem[{\citenamefont{Cao et~al.}(1998)\citenamefont{Cao, Linden, Farnebo,
  Cao, Eriksson, Kumar, Qi, Claesson-Welsh, and Alitalo}}]{Cao1998}
\bibinfo{author}{\bibfnamefont{Y.}~\bibnamefont{Cao}},
  \bibinfo{author}{\bibfnamefont{P.}~\bibnamefont{Linden}},
  \bibinfo{author}{\bibfnamefont{J.}~\bibnamefont{Farnebo}},
  \bibinfo{author}{\bibfnamefont{R.}~\bibnamefont{Cao}},
  \bibinfo{author}{\bibfnamefont{A.}~\bibnamefont{Eriksson}},
  \bibinfo{author}{\bibfnamefont{V.}~\bibnamefont{Kumar}},
  \bibinfo{author}{\bibfnamefont{J.~H.} \bibnamefont{Qi}},
  \bibinfo{author}{\bibfnamefont{L.}~\bibnamefont{Claesson-Welsh}},
  \bibnamefont{and} \bibinfo{author}{\bibfnamefont{K.}~\bibnamefont{Alitalo}},
  \bibinfo{journal}{P. Natl. Acad. Sci. U.S.A.} \textbf{\bibinfo{volume}{95}},
  \bibinfo{pages}{14389} (\bibinfo{year}{1998}).

\bibitem[{\citenamefont{Drake et~al.}(2000)\citenamefont{Drake, LaRue, Ferrara,
  and Little}}]{Drake2000}
\bibinfo{author}{\bibfnamefont{C.~J.} \bibnamefont{Drake}},
  \bibinfo{author}{\bibfnamefont{A.}~\bibnamefont{LaRue}},
  \bibinfo{author}{\bibfnamefont{N.}~\bibnamefont{Ferrara}}, \bibnamefont{and}
  \bibinfo{author}{\bibfnamefont{C.~D.} \bibnamefont{Little}},
  \bibinfo{journal}{Dev. Biol.} \textbf{\bibinfo{volume}{224}},
  \bibinfo{pages}{178} (\bibinfo{year}{2000}).

\bibitem[{\citenamefont{Parsa et~al.}(2011)\citenamefont{Parsa, Upadhyay, and
  Sia}}]{Parsa22032011}
\bibinfo{author}{\bibfnamefont{H.}~\bibnamefont{Parsa}},
  \bibinfo{author}{\bibfnamefont{R.}~\bibnamefont{Upadhyay}}, \bibnamefont{and}
  \bibinfo{author}{\bibfnamefont{S.~K.} \bibnamefont{Sia}},
  \bibinfo{journal}{Proc. Natl. Acad. Sci. U.S.A.}
  \textbf{\bibinfo{volume}{108}}, \bibinfo{pages}{5133} (\bibinfo{year}{2011}).

\bibitem[{\citenamefont{Gamba et~al.}(2003)\citenamefont{Gamba, Ambrosi,
  Coniglio, {De Candia}, {Di Talia}, Giraudo, Serini, Preziosi, and
  Bussolino}}]{Gamba2003}
\bibinfo{author}{\bibfnamefont{A.}~\bibnamefont{Gamba}},
  \bibinfo{author}{\bibfnamefont{D.}~\bibnamefont{Ambrosi}},
  \bibinfo{author}{\bibfnamefont{A.}~\bibnamefont{Coniglio}},
  \bibinfo{author}{\bibfnamefont{A.}~\bibnamefont{{De Candia}}},
  \bibinfo{author}{\bibfnamefont{S.}~\bibnamefont{{Di Talia}}},
  \bibinfo{author}{\bibfnamefont{E.}~\bibnamefont{Giraudo}},
  \bibinfo{author}{\bibfnamefont{G.}~\bibnamefont{Serini}},
  \bibinfo{author}{\bibfnamefont{L.}~\bibnamefont{Preziosi}}, \bibnamefont{and}
  \bibinfo{author}{\bibfnamefont{F.}~\bibnamefont{Bussolino}},
  \bibinfo{journal}{Phys. Rev. Lett.} \textbf{\bibinfo{volume}{90}},
  \bibinfo{pages}{118101} (\bibinfo{year}{2003}).

\bibitem[{\citenamefont{Serini et~al.}(2003)\citenamefont{Serini, Ambrosi,
  Giraudo, Gamba, Preziosi, and Bussolino}}]{Serini2003}
\bibinfo{author}{\bibfnamefont{G.}~\bibnamefont{Serini}},
  \bibinfo{author}{\bibfnamefont{D.}~\bibnamefont{Ambrosi}},
  \bibinfo{author}{\bibfnamefont{E.}~\bibnamefont{Giraudo}},
  \bibinfo{author}{\bibfnamefont{A.}~\bibnamefont{Gamba}},
  \bibinfo{author}{\bibfnamefont{L.}~\bibnamefont{Preziosi}}, \bibnamefont{and}
  \bibinfo{author}{\bibfnamefont{F.}~\bibnamefont{Bussolino}},
  \bibinfo{journal}{EMBO J.} \textbf{\bibinfo{volume}{22}},
  \bibinfo{pages}{1771} (\bibinfo{year}{2003}).

\bibitem[{\citenamefont{Ambrosi et~al.}(2004)\citenamefont{Ambrosi, Gamba, and
  Serini}}]{Ambrosi2004}
\bibinfo{author}{\bibfnamefont{D.}~\bibnamefont{Ambrosi}},
  \bibinfo{author}{\bibfnamefont{A.}~\bibnamefont{Gamba}}, \bibnamefont{and}
  \bibinfo{author}{\bibfnamefont{G.}~\bibnamefont{Serini}},
  \bibinfo{journal}{Bull. Math. Biol.} \textbf{\bibinfo{volume}{66}},
  \bibinfo{pages}{1851} (\bibinfo{year}{2004}).

\bibitem[{\citenamefont{Merks et~al.}(2004)\citenamefont{Merks, Newman, and
  Glazier}}]{Merks2004}
\bibinfo{author}{\bibfnamefont{R.~M.~H.} \bibnamefont{Merks}},
  \bibinfo{author}{\bibfnamefont{S.~A.} \bibnamefont{Newman}},
  \bibnamefont{and} \bibinfo{author}{\bibfnamefont{J.~A.}
  \bibnamefont{Glazier}}, \bibinfo{journal}{Lect. Notes. Comput. Sc.} pp.
  \bibinfo{pages}{425--434} (\bibinfo{year}{2004}).

\bibitem[{\citenamefont{Merks et~al.}(2006)\citenamefont{Merks, Brodsky,
  Goligorksy, Newman, and Glazier}}]{Merks2006b}
\bibinfo{author}{\bibfnamefont{R.~M.~H.} \bibnamefont{Merks}},
  \bibinfo{author}{\bibfnamefont{S.~V.} \bibnamefont{Brodsky}},
  \bibinfo{author}{\bibfnamefont{M.~S.} \bibnamefont{Goligorksy}},
  \bibinfo{author}{\bibfnamefont{S.~A.} \bibnamefont{Newman}},
  \bibnamefont{and} \bibinfo{author}{\bibfnamefont{J.~A.}
  \bibnamefont{Glazier}}, \bibinfo{journal}{Dev. Biol.}
  \textbf{\bibinfo{volume}{289}}, \bibinfo{pages}{44} (\bibinfo{year}{2006}).

\bibitem[{\citenamefont{Merks and Glazier}(2006)}]{Merks2006a}
\bibinfo{author}{\bibfnamefont{R.~M.~H.} \bibnamefont{Merks}} \bibnamefont{and}
  \bibinfo{author}{\bibfnamefont{J.~A.} \bibnamefont{Glazier}},
  \bibinfo{journal}{Nonlinearity} \textbf{\bibinfo{volume}{19}}
  (\bibinfo{year}{2006}).

\bibitem[{\citenamefont{Merks et~al.}(2008)\citenamefont{Merks, Perryn,
  Shirinifard, and Glazier}}]{Merks2008}
\bibinfo{author}{\bibfnamefont{R.~M.~H.} \bibnamefont{Merks}},
  \bibinfo{author}{\bibfnamefont{E.~D.} \bibnamefont{Perryn}},
  \bibinfo{author}{\bibfnamefont{A.}~\bibnamefont{Shirinifard}},
  \bibnamefont{and} \bibinfo{author}{\bibfnamefont{J.~A.}
  \bibnamefont{Glazier}}, \bibinfo{journal}{PLoS Comput. Biol.}
  \textbf{\bibinfo{volume}{4}}, \bibinfo{pages}{e1000163}
  (\bibinfo{year}{2008}).

\bibitem[{\citenamefont{K\"{o}hn-Luque
  et~al.}(2011)\citenamefont{K\"{o}hn-Luque, de~Back, Starruss, Mattiotti,
  Deutsch, P\'{e}rez-Pomares, and Herrero}}]{Kohn-Luque2011}
\bibinfo{author}{\bibfnamefont{A.}~\bibnamefont{K\"{o}hn-Luque}},
  \bibinfo{author}{\bibfnamefont{W.}~\bibnamefont{de~Back}},
  \bibinfo{author}{\bibfnamefont{J.}~\bibnamefont{Starruss}},
  \bibinfo{author}{\bibfnamefont{A.}~\bibnamefont{Mattiotti}},
  \bibinfo{author}{\bibfnamefont{A.}~\bibnamefont{Deutsch}},
  \bibinfo{author}{\bibfnamefont{J.~M.} \bibnamefont{P\'{e}rez-Pomares}},
  \bibnamefont{and} \bibinfo{author}{\bibfnamefont{M.~a.}
  \bibnamefont{Herrero}}, \bibinfo{journal}{PloS ONE}
  \textbf{\bibinfo{volume}{6}}, \bibinfo{pages}{e24175} (\bibinfo{year}{2011}).

\bibitem[{\citenamefont{Scianna et~al.}(2011)\citenamefont{Scianna, Munaron,
  and Preziosi}}]{Scianna2011}
\bibinfo{author}{\bibfnamefont{M.}~\bibnamefont{Scianna}},
  \bibinfo{author}{\bibfnamefont{L.}~\bibnamefont{Munaron}}, \bibnamefont{and}
  \bibinfo{author}{\bibfnamefont{L.}~\bibnamefont{Preziosi}},
  \bibinfo{journal}{Prog. Biophys. Mol. Bio.} pp. \bibinfo{pages}{1--20}
  (\bibinfo{year}{2011}).

\bibitem[{\citenamefont{Manoussaki et~al.}(1996)\citenamefont{Manoussaki,
  Lubkin, Vemon, and Murray}}]{Manoussaki1996}
\bibinfo{author}{\bibfnamefont{D.}~\bibnamefont{Manoussaki}},
  \bibinfo{author}{\bibfnamefont{S.~R.} \bibnamefont{Lubkin}},
  \bibinfo{author}{\bibfnamefont{R.~B.} \bibnamefont{Vemon}}, \bibnamefont{and}
  \bibinfo{author}{\bibfnamefont{J.~D.} \bibnamefont{Murray}},
  \bibinfo{journal}{Acta Biotheor.} \textbf{\bibinfo{volume}{44}},
  \bibinfo{pages}{271} (\bibinfo{year}{1996}).

\bibitem[{\citenamefont{Manoussaki}(2002)}]{Manoussaki2002}
\bibinfo{author}{\bibfnamefont{D.}~\bibnamefont{Manoussaki}}, in
  \emph{\bibinfo{booktitle}{ESAIM: Proceedings}} (\bibinfo{year}{2002}),
  vol.~\bibinfo{volume}{12}, pp. \bibinfo{pages}{108--114}.

\bibitem[{\citenamefont{Murray}(2003)}]{Murray2003}
\bibinfo{author}{\bibfnamefont{J.~D.} \bibnamefont{Murray}},
  \bibinfo{journal}{C. R. Biol.} \textbf{\bibinfo{volume}{326}},
  \bibinfo{pages}{239} (\bibinfo{year}{2003}).

\bibitem[{\citenamefont{Tranqui and Tracqui}(2000)}]{Tranqui2000}
\bibinfo{author}{\bibfnamefont{L.}~\bibnamefont{Tranqui}} \bibnamefont{and}
  \bibinfo{author}{\bibfnamefont{P.}~\bibnamefont{Tracqui}},
  \bibinfo{journal}{C. R. Acad. Sci. III-Vie.} \textbf{\bibinfo{volume}{323}},
  \bibinfo{pages}{31} (\bibinfo{year}{2000}).

\bibitem[{\citenamefont{Namy et~al.}(2004)\citenamefont{Namy, Ohayon, and
  Tracqui}}]{Namy2004a}
\bibinfo{author}{\bibfnamefont{P.}~\bibnamefont{Namy}},
  \bibinfo{author}{\bibfnamefont{J.}~\bibnamefont{Ohayon}}, \bibnamefont{and}
  \bibinfo{author}{\bibfnamefont{P.}~\bibnamefont{Tracqui}},
  \bibinfo{journal}{J. Theor. Biol.} \textbf{\bibinfo{volume}{227}},
  \bibinfo{pages}{103} (\bibinfo{year}{2004}).

\bibitem[{\citenamefont{Tracqui et~al.}(2005)\citenamefont{Tracqui, Namy, and
  Ohayon}}]{Tracqui2002}
\bibinfo{author}{\bibfnamefont{P.}~\bibnamefont{Tracqui}},
  \bibinfo{author}{\bibfnamefont{P.}~\bibnamefont{Namy}}, \bibnamefont{and}
  \bibinfo{author}{\bibfnamefont{J.}~\bibnamefont{Ohayon}},
  \bibinfo{journal}{J. Biol. Phys. Chem.} \textbf{\bibinfo{volume}{5}},
  \bibinfo{pages}{57} (\bibinfo{year}{2005}).

\bibitem[{\citenamefont{Szab\'{o} et~al.}(2007)\citenamefont{Szab\'{o}, Perryn,
  and Czir\'{o}k}}]{Szabo2007}
\bibinfo{author}{\bibfnamefont{A.}~\bibnamefont{Szab\'{o}}},
  \bibinfo{author}{\bibfnamefont{E.~D.} \bibnamefont{Perryn}},
  \bibnamefont{and}
  \bibinfo{author}{\bibfnamefont{A.}~\bibnamefont{Czir\'{o}k}},
  \bibinfo{journal}{Phys. Rev. Lett.} \textbf{\bibinfo{volume}{98}},
  \bibinfo{pages}{038102} (\bibinfo{year}{2007}).

\bibitem[{\citenamefont{Szab\'{o} et~al.}(2008)\citenamefont{Szab\'{o}, Mehes,
  Kosa, and Czir\'{o}k}}]{Szabo2008}
\bibinfo{author}{\bibfnamefont{A.}~\bibnamefont{Szab\'{o}}},
  \bibinfo{author}{\bibfnamefont{E.}~\bibnamefont{Mehes}},
  \bibinfo{author}{\bibfnamefont{E.}~\bibnamefont{Kosa}}, \bibnamefont{and}
  \bibinfo{author}{\bibfnamefont{A.}~\bibnamefont{Czir\'{o}k}},
  \bibinfo{journal}{Biophys. J.} \textbf{\bibinfo{volume}{95}},
  \bibinfo{pages}{2702} (\bibinfo{year}{2008}).

\bibitem[{\citenamefont{Graner and Glazier}(1992)}]{Graner1992}
\bibinfo{author}{\bibfnamefont{F.}~\bibnamefont{Graner}} \bibnamefont{and}
  \bibinfo{author}{\bibfnamefont{J.~A.} \bibnamefont{Glazier}},
  \bibinfo{journal}{Phys. Rev. Lett.} \textbf{\bibinfo{volume}{69}},
  \bibinfo{pages}{2013} (\bibinfo{year}{1992}).

\bibitem[{\citenamefont{Glazier and Graner}(1993)}]{Glazier1993a}
\bibinfo{author}{\bibfnamefont{J.~A.} \bibnamefont{Glazier}} \bibnamefont{and}
  \bibinfo{author}{\bibfnamefont{F.}~\bibnamefont{Graner}},
  \bibinfo{journal}{Phys. Rev. E} \textbf{\bibinfo{volume}{47}},
  \bibinfo{pages}{2128} (\bibinfo{year}{1993}).

\bibitem[{\citenamefont{Glazier et~al.}(2008)\citenamefont{Glazier, Zhang,
  Swat, Zaitlen, and Schnell}}]{Glazier2008}
\bibinfo{author}{\bibfnamefont{J.~A.} \bibnamefont{Glazier}},
  \bibinfo{author}{\bibfnamefont{Y.}~\bibnamefont{Zhang}},
  \bibinfo{author}{\bibfnamefont{M.~H.} \bibnamefont{Swat}},
  \bibinfo{author}{\bibfnamefont{B.}~\bibnamefont{Zaitlen}}, \bibnamefont{and}
  \bibinfo{author}{\bibfnamefont{S.}~\bibnamefont{Schnell}},
  \bibinfo{journal}{Curr. Top. Dev. Biol.} \textbf{\bibinfo{volume}{81}}
  (\bibinfo{year}{2008}).

\bibitem[{\citenamefont{Hester et~al.}(2011)\citenamefont{Hester, Belmonte,
  Gens, Clendenon, and Glazier}}]{Hester2011}
\bibinfo{author}{\bibfnamefont{S.~D.} \bibnamefont{Hester}},
  \bibinfo{author}{\bibfnamefont{J.~M.} \bibnamefont{Belmonte}},
  \bibinfo{author}{\bibfnamefont{J.~S.} \bibnamefont{Gens}},
  \bibinfo{author}{\bibfnamefont{S.~G.} \bibnamefont{Clendenon}},
  \bibnamefont{and} \bibinfo{author}{\bibfnamefont{J.~a.}
  \bibnamefont{Glazier}}, \bibinfo{journal}{PLoS Comput. Biol.}
  \textbf{\bibinfo{volume}{7}}, \bibinfo{pages}{e1002155}
  (\bibinfo{year}{2011}).

\bibitem[{\citenamefont{Zajac et~al.}(2000)\citenamefont{Zajac, Jones, and
  Glazier}}]{Zajac2000a}
\bibinfo{author}{\bibfnamefont{M.}~\bibnamefont{Zajac}},
  \bibinfo{author}{\bibfnamefont{G.~L.} \bibnamefont{Jones}}, \bibnamefont{and}
  \bibinfo{author}{\bibfnamefont{J.~A.} \bibnamefont{Glazier}},
  \bibinfo{journal}{Phys. Rev. Lett.} \textbf{\bibinfo{volume}{85}},
  \bibinfo{pages}{2022} (\bibinfo{year}{2000}).

\bibitem[{\citenamefont{K\"{a}fer et~al.}(2007)\citenamefont{K\"{a}fer,
  Hayashi, Mar\'{e}e, Carthew, and Graner}}]{Kafer2007}
\bibinfo{author}{\bibfnamefont{J.}~\bibnamefont{K\"{a}fer}},
  \bibinfo{author}{\bibfnamefont{T.}~\bibnamefont{Hayashi}},
  \bibinfo{author}{\bibfnamefont{A.~F.~M.} \bibnamefont{Mar\'{e}e}},
  \bibinfo{author}{\bibfnamefont{R.~W.} \bibnamefont{Carthew}},
  \bibnamefont{and} \bibinfo{author}{\bibfnamefont{F.}~\bibnamefont{Graner}},
  \bibinfo{journal}{P. Natl. Acad. Sci. U.S.A.} \textbf{\bibinfo{volume}{104}},
  \bibinfo{pages}{18549} (\bibinfo{year}{2007}).

\bibitem[{\citenamefont{Savill and Hogeweg}(1997)}]{Savill1996}
\bibinfo{author}{\bibfnamefont{N.}~\bibnamefont{Savill}} \bibnamefont{and}
  \bibinfo{author}{\bibfnamefont{P.}~\bibnamefont{Hogeweg}},
  \bibinfo{journal}{J. Theor. Biol.} \textbf{\bibinfo{volume}{184}},
  \bibinfo{pages}{229} (\bibinfo{year}{1997}).

\bibitem[{\citenamefont{{De Gennes} and Prost}(1993)}]{DeGennes1993}
\bibinfo{author}{\bibfnamefont{P.~G.} \bibnamefont{{De Gennes}}}
  \bibnamefont{and} \bibinfo{author}{\bibfnamefont{J.}~\bibnamefont{Prost}},
  \emph{\bibinfo{title}{{The physics of liquid crystals}}}
  (\bibinfo{publisher}{Oxford University Press}, \bibinfo{year}{1993}),
  \bibinfo{edition}{2nd} ed.

\bibitem[{\citenamefont{Foffi et~al.}(2005)\citenamefont{Foffi, {De Michele},
  Sciortino, and Tartaglia}}]{Foffi2005}
\bibinfo{author}{\bibfnamefont{G.}~\bibnamefont{Foffi}},
  \bibinfo{author}{\bibfnamefont{C.}~\bibnamefont{{De Michele}}},
  \bibinfo{author}{\bibfnamefont{F.}~\bibnamefont{Sciortino}},
  \bibnamefont{and}
  \bibinfo{author}{\bibfnamefont{P.}~\bibnamefont{Tartaglia}},
  \bibinfo{journal}{J. Chem. Phys.} \textbf{\bibinfo{volume}{122}},
  \bibinfo{pages}{224903} (\bibinfo{year}{2005}).

\bibitem[{\citenamefont{Solomon and Spicer}(2010)}]{Solomon2010}
\bibinfo{author}{\bibfnamefont{M.~J.} \bibnamefont{Solomon}} \bibnamefont{and}
  \bibinfo{author}{\bibfnamefont{P.~T.} \bibnamefont{Spicer}},
  \bibinfo{journal}{Soft Matter} \textbf{\bibinfo{volume}{6}},
  \bibinfo{pages}{1391} (\bibinfo{year}{2010}).

\bibitem[{\citenamefont{Bernal and Fankuchen}(1941)}]{Bernal:1941to}
\bibinfo{author}{\bibfnamefont{J.~D.} \bibnamefont{Bernal}} \bibnamefont{and}
  \bibinfo{author}{\bibfnamefont{I.}~\bibnamefont{Fankuchen}},
  \bibinfo{journal}{J Gen Physiol} \textbf{\bibinfo{volume}{25}},
  \bibinfo{pages}{111} (\bibinfo{year}{1941}).

\bibitem[{\citenamefont{van Bruggen and
  Lekkerkerker}(2002)}]{vanBruggen:2002hf}
\bibinfo{author}{\bibfnamefont{M.~P.~B.} \bibnamefont{van Bruggen}}
  \bibnamefont{and} \bibinfo{author}{\bibfnamefont{H.~N.~W.}
  \bibnamefont{Lekkerkerker}}, \bibinfo{journal}{Langmuir}
  \textbf{\bibinfo{volume}{18}}, \bibinfo{pages}{7141} (\bibinfo{year}{2002}).

\bibitem[{\citenamefont{Wilkins et~al.}(2009)\citenamefont{Wilkins, Spicer, and
  Solomon}}]{Wilkins:2009ht}
\bibinfo{author}{\bibfnamefont{G.~M.~H.} \bibnamefont{Wilkins}},
  \bibinfo{author}{\bibfnamefont{P.~T.} \bibnamefont{Spicer}},
  \bibnamefont{and} \bibinfo{author}{\bibfnamefont{M.~J.}
  \bibnamefont{Solomon}}, \bibinfo{journal}{Langmuir}
  \textbf{\bibinfo{volume}{25}}, \bibinfo{pages}{8951} (\bibinfo{year}{2009}).

\bibitem[{\citenamefont{Swat et~al.}(2012)\citenamefont{Swat, Thomas, and
  Belmonte}}]{Swat2012}
\bibinfo{author}{\bibfnamefont{M.~H.} \bibnamefont{Swat}},
  \bibinfo{author}{\bibfnamefont{G.~L.} \bibnamefont{Thomas}},
  \bibnamefont{and} \bibinfo{author}{\bibfnamefont{J.~M.}
  \bibnamefont{Belmonte}}, \bibinfo{journal}{Method. Cell. Biol.}
  \textbf{\bibinfo{volume}{110}}, \bibinfo{pages}{325} (\bibinfo{year}{2012}).

\end{thebibliography}

\end{document}